\documentclass[preprint,showpacs,preprintnumbers,nofootinbib,amsmath,amssymb,aps,prc,longbibliography]{revtex4-1}
\usepackage{graphicx}
\usepackage{dcolumn}
\usepackage{bm}
\usepackage{epstopdf}
\usepackage{xcolor}

\begin{document}
\title{Improved treatment of blocking effect at finite temperature}
\author{N. Quang Hung$^{1}$}
 \email{nqhungdtu@gmail.com}
 \author{N. Dinh Dang$^{2,3}$}
  \email{dang@riken.jp}
  \author{L.T. Quynh Huong$^{1,4,5}$}
 \affiliation{1) Institute of Research and Development, Duy Tan University, K7/25 Quang Trung, Danang City, Vietnam\\
2) Quantum Hadron Physics Laboratory, RIKEN Nishina Center
for Accelerator-Based Science,
2-1 Hirosawa, Wako City, 351-0198 Saitama, Japan\\
3) Institute for Nuclear Science and Technique, Hanoi, Vietnam\\
4) Department of Natural Science and Technology, University of Khanh Hoa, Nha Trang City, Khanh Hoa Province, Vietnam\\
5) Faculty of Physics and Engineering Physics, Ho Chi Minh University of Science, Ho Chi Minh City, Vietnam}

\date{\today}
\begin{abstract}
The blocking effect caused by the odd particle on the pairing properties of systems with an odd number of fermions at finite temperature interacting via the monopole pairing force is studied within several approximations. The results are compared with the predictions obtained by using the exact solutions of the pairing Hamiltonian.
The comparison favors the approximation with the odd particle occupying the top level, which is the closest to the Fermi surface and whose occupation number decreases with increasing temperature.

\end{abstract}

\pacs{21.60.-n, 21.60.Jz, 24.60.-k, 24.10.Pa}
\keywords{Suggested keywords}
\maketitle
\section{Introduction}
\label{Intro}
The BCS theory is a most popular approximation for treating pairing interaction between fermions such as nucleons in atomic nuclei. In the systems with an  odd number of particles at zero temperature ($T=$ 0), which are referred to hereafter as odd systems, the Pauli principle prevents the level $k$, which is occupied by the odd particle, from the pair scattering process due to pairing correlations. This level $k$ remains unpaired with the occupation number equal to 1 and, consequently, is excluded from the BCS equation for the pairing gap, whereas the equation for the particle number is solved only for the even core. This approximation is traditionally referred to as the ``blocked BCS"~\cite{Wahlborn,Ring}.  Although the odd particle can occupy any level $k$ above the Fermi surface, at $T=$ 0 the lowest level $k_0$,  which is located just above the Fermi surface, corresponds to the ground state of the odd system. Any physical observable such as the ground state energy, pairing gap, etc., is obtained by minimizing the pairing Hamiltonian, which is averaged over the ground state.  

At finite temperature ($T\neq$ 0), the average over the ground state is replaced by that over a statistical ensemble such as the grand canonical ensemble (GCE), where the energy and particle number of the system fluctuate in contact with the heat bath, or canonical ensemble (CE), where the system with a fixed number of particle exchanges its energy with the heat bath. Several approximations were proposed to describe the properties of odd systems at finite temperature.

The first type of approximation takes into account the blocking effect only at zero temperature ($T=$ 0). As this reduces the pairing gap at $T=$ 0, the approximation is simply equivalent to a shift in the excitation energy. The second type of approximation is the so-called fixed quasiparticle-number approach, which was proposed by Ignatyuk and Sokolov~\cite{Ignatyuk1} to describe the odd system by considering only states consisting of an odd number of quasiparticles. Because this approach is constructed on the basis of the BCS level density calculations with a fixed number of quasiparticles ~\cite{Ignatyuk2,Moretto}, it does not deal explicitly with the odd BCS equations at finite $T$, and hence, it will not be discussed in the present paper.

The third type of approximation, proposed by Maino \textit{et al.} \cite{Maino}, is a direct extension of the blocked BCS to finite temperature $T$.  This approach assumes that, at $T\neq$ 0, the odd particle occupies the level $k$ with the unity occupation number ($q_k =$ 1) as at $T=$ 0, but $k$ can be any level above the Fermi surface, not only $k=k_0$. The grand partition function is calculated with the trace taken over all possible positions of the blocked level $k$. The approach then takes the average of any observable as the sum of all its average values taken for each blocked level $k$ with the statistical weight as the ratio of the grand partition function for the blocked level $k$ to the total grand partition function. For simplicity, we refer to this approach as ``Maino's" hereafter. 

The assumption of the unity occupation probability of the odd level in the Maino's approach means that temperatures have no effect on it, whereas the occupation numbers of all the levels in the pairing core are obtained by averaging over the GCE, and, they are therefore temperature-dependent.  An obvious inconsistency of such assumption is that it fails to reproduce the zero-pairing limit of the equation for the particle number at finite temperature, where all the single-particle occupation numbers should follow the Fermi-Dirac distribution of noninteracting fermions. 

In the present paper we propose to improve the blocked BCS to include the temperature dependence of the occupation number $q_k$ for the level $k$, which is occupied by the odd particle. We refer to this approach as ``$q_k$-blocked BCS". The justification of this modification is given by analyzing the exact solutions of the pairing problem, which are obtained by diagonalizing the pairing Hamiltonian, and used to construct the CE partition function. Based on the $q_k$-blocked BCS, we propose an approximation, the average-blocked BCS, which reproduces the results of Maino's approach, but does not contain the inconsistency of the latter in recovering the zero-temperature limit (to be explained in the paper). Moreover, by comparing with the exact solutions of the pairing problem, we will demonstrate that the $q_{k_0}$-blocked BCS with $k=k_0$ is the most appropriate approximation for the description of the pairing properties of odd systems at finite temperature. An extension of the $q_{k_0}$-blocked BCS, which takes into account the quasiparticle number fluctuation (QNF), is also carried out. This extension is based on the approach called FTBCS1~\cite{FTBCS1} for even systems, where the QNF smoothes out the sharp phase transition between the superfluid phase to the normal one at the critical temperature $T_c\simeq$ 0.57 $\Delta(0)$ with $\Delta(0)$ being the pairing gap at $T=$ 0.

The paper is organized as follows. The $q_k$-blocked BCS and average-blocked BCS are presented in Sec. \ref{formalism}, where the Maino's approximation and exact solutions of the pairing problem, which are extended to $T\neq$ 0, are also summarized. The results of numerical calculations are analyzed in Sec. \ref{results}. The paper is summarized in the last section, where conclusions are drawn.
\section{Formalism}
\label{formalism}
\subsection{Pairing Hamiltonian and the finite-temperature BCS equations for even nuclei}
We employ the standard pairing Hamiltonian of a system, which consists of $N$ particles
interacting via a monopole pairing force with the constant parameter $G$, namely,
\begin{equation}
\hat{H}=\sum_k\epsilon_k(a_{+k}^{\dagger}a_{+k}+a_{-k}^{\dagger}a_{-k})-G\sum_{kk'}{a_k^{\dagger}
a_{-k}^{\dagger}a_{-k'}a_{k'}}
\label{Heven}
\end{equation}
with $a_{\pm k}^{\dagger}(a_{\pm k})$ being the creation (annihilation) operators of a particle (neutron or proton) with angular momentum $k$, projection $\pm m_k$, and energy $\epsilon_k$. This means this level $k$ is doubly degenerate, that is, it can be occupied at most by two particles in the states $|k,m_k\rangle$ and $|k,-m_k\rangle$ with the same energy $\epsilon_k$. The particle-number operator $\hat{N}$ is given as
\begin{equation}
\hat{N}=\sum_k(a_{+k}^{\dagger}a_{+k} + a_{-k}^{\dagger}a_{-k})~. \label{N}
\end{equation}

After the Bogoliubov transformation from the particle operators, $a_k^\dagger$ and  $a_k$, to the quasiparticle ones, $\alpha_k^\dagger$ and $\alpha_k$,
\begin{equation}
a_k^\dagger = u_k\alpha_k^\dagger + v_k\alpha_{-k}~, \hspace{5mm}
a_{-k}=u_k\alpha_{-k}  - v_k\alpha_k^\dagger ~, \label{Bogo}
\end{equation}
the Hamiltonian \eqref{Heven} is transformed into the quasiparticle one $\cal H$, whose explicit form is given, e.g., in Ref. \cite{FTBCS1}.

By applying the Bogoliubov transformation and the average in the GCE to the operator (\ref{N}), one obtains the average particle number $N$ in the even system as
\begin{equation}
N \equiv\langle\hat{N}\rangle =2\sum_k[ n_k u^2_k+ (1-n_k) v_k^2]~,
\label{Neven} 
\end{equation} 
where $\langle\hat{\cal O}\rangle$ denotes the GCE average 
\begin{equation}
\langle\hat{\cal O}\rangle = \frac{\texttt{Tr}[\hat{\cal O}e^{-\beta(\hat{H}-\lambda\hat{N})}]}{\texttt{Tr}[e^{-\beta(\hat{H}-\lambda\hat{N})}]}~,\hspace{5mm} \beta = \frac{1}{T}~,  
\label{GCE}
\end{equation}
$N$ is an even number, and $n_k$ is the quasiparticle occupation number
\begin{equation}
n_k= \frac{1}{e^{\beta E_k} +1}
\label{nk}
\end{equation}
with the chemical potential $\lambda$ and the quasiparticle energy 
\begin{equation}
E_k = \sqrt{(\epsilon_k - \lambda)^2 + \Delta^2}~,
\label{Ek}
\end{equation}
where the self-energy correction term $-Gv_k^2$ is neglected in $\epsilon_k-\lambda$ for simplicity, because it has only a small effect at very low temperature in the present model (see Figs. 2 in Refs. \cite{DangPRC76,HungPRC81} and discussions therein). This does not affect the temperature dependence of all thermodynamic quantities considered in the present paper. In the realistic calculations, the single-particle energies deduced from the experimental data are often used, including this self-energy correction term.

The pairing gap $\Delta$ is found by minimizing the average value $\langle \hat{H} - \lambda \hat{N}\rangle$, resulting in the BCS gap equation
\begin{equation}
\Delta = G\sum_k(1-2n_k)u_kv_k~.
\label{BCSgap}
\end{equation} 

The coefficients $u_k$ and $v_k$ are expressed in terms of the chemical potential, the pairing gap $\Delta$, and the single-particle energy $\epsilon_k$ as
\begin{equation}
u_k=\sqrt{\frac{1}{2}\bigg(1 + \frac{\epsilon_k - \lambda}{E_k}\bigg)}~,\hspace{3mm} v_k=\sqrt{\frac{1}{2}\bigg(1 - \frac{\epsilon_k - \lambda}{E_k}\bigg)}~.
\label{uv}
\end{equation}

The set of two equations (\ref{Neven}) and (\ref{BCSgap}) forms the so-called finite-temperature BCS (FT-BCS) equations to determine the chemical potential $\lambda$ and the pairing gap $\Delta$ for a system with even particle number $N$ interacting via 
the pairing force with parameter $G$ at a given temperature $T$.

The thermodynamic quantities such as the total energy $E$, the heat capacity $C$, and the quasiparticle entropy $S$ are calculated in the standard way as
\begin{equation}
E(T)\equiv\langle \hat{H}\rangle= \sum_{k}\epsilon_k[u^2_kn_k+(1-n_k)v_k^2]~,\hspace{3mm} C = \frac{dE}{dT}~,
\label{EC}
\end{equation}
\begin{equation}
S = -2\sum_k[n_k\ln n_k +(1-n_k)\ln(1-n_k)]~.
\label{S}
\end{equation} 

The excitation energy $E^*$ is given as the difference between $E(T)$ and its ground state value $E(0)$, namely,
\begin{equation}
E^*(T) = E(T) - E(0)~.
\label{E*}
\end{equation}
\subsection{Maino's approach}
\label{Maino}
In the odd system, where the odd particle occupies the level $k$, this level is  blocked from pair scattering and excluded from the FT-BCS equations (\ref{Neven}) and (\ref{BCSgap}). Assuming the unity occupation number $q_k =$ 1 of the odd level $k$ as that at $T=$0, the blocked-BCS equations at finite $T$ are written in the Maino's approach as
\begin{equation}
\Delta^{(k)} = G\sum_{k'\neq k}(1-2n_{k'})u_{k'}v_{k'}~,
\label{oddBCSgap}
\end{equation} 
\begin{equation}
N^{(k)} = 1+ 2\sum_{k'\neq k}[n_{k'}u_{k'}^2 + (1-n_{k'}) v_{k'}^2]~,
\label{Nodd} 
\end{equation}
where the total particle number $N$ of the system is an odd number. The superscript $(k)$ denotes the level $k$ that is occupied by the odd particle and excluded from the pairing even core.

To calculate the thermodynamic quantities, it is further assumed that $k$ can be any level above the chemical potential $\lambda$, and the grand partition function $Z(\beta,\lambda)$ is constructed as a sum of all partition functions $Z^{(k)}$ of the systems with the blocked level $k$, namely,
\begin{equation}
Z(\beta,\lambda) =\texttt{Tr}\{e^{-\beta[\hat{H}^{(k)} - \lambda\hat{N}^{(k)}]}\}=\sum_{k}Z^{(k)}(\beta,\lambda)~.
\label{Z}
\end{equation}

In Eq. (\ref{Z}) $\hat{H}^{(k)}$ is the Hamiltonian of the odd-particle-number system, where the level $k$ is occupied by the odd particle:
\begin{eqnarray}
&&\hat{H}^{(k)}=\epsilon_k(a_{+k}^{\dagger}a_{+k}+a_{-k}^{\dagger}a_{-k})+\hat{H}'~, \nonumber \\ 
&&\hat{H'}=\sum_{k'\neq k}\epsilon_{k'}(a_{+{k'}}^{\dagger}a_{+{k'}}+a_{-{k'}}^{\dagger}a_{-{k'}}) - G\sum_{k',k''\neq k}{a_{k'}^{\dagger}a_{-{k'}}^{\dagger}a_{-k''}a_{k''}}~.
\label{Hodd}
\end{eqnarray}

The average value $\langle\hat{\cal O}^{(k)}\rangle$ of an operator $\hat{\cal O}^{(k)}$ over all possible configurations of the system is given by
\begin{equation}
\langle\hat{\cal O}^{(k)}\rangle =\sum_{k}R^{(k)}{\cal O}^{(k)}~,\hspace{5mm} R^{(k)}=\frac{Z^{(k)}}{Z}~.
\label{Ok}
\end{equation}

The total energy, heat capacity and quasiparticle entropy are then calculated as
\begin{equation}
E^{Maino} = \sum_{k}R^{(k)}(\epsilon_k + \langle\hat{H}'\rangle)~,\hspace{5mm} C^{Maino} = \frac{dE_{Maino}}{dT}~,
\label{ECMaino}
\end{equation}
\begin{equation}
S^{Maino} = \sum_k R^{(k)}S^{(k)}~, \hspace{5mm}
S^{(k)} = -2\sum_{k'\neq k}[n_{k'}\ln n_{k'}+(1-n_{k'})\ln(1-n_{k'})].
\label{SMaino}
\end{equation} 

To restore the zero-temperature limit where $k=k_0$, which is the first level above the ground-state chemical potential $\lambda(T=0)$, the following requirement is imposed within the Maino's approach:
\begin{equation}
\lim_{\beta\rightarrow\infty}R^{(k)} = \delta_{kk_0}~.
\label{limR}
\end{equation}

\subsection{$q_k$-blocked BCS and average-blocked BCS} 
\label{q-oddBCS}
An obvious shortcoming of the assumption $q_k=$ 1 in the blocked-BCS Eq. (\ref{Nodd}) is that it fails to reproduce the zero-pairing limit at $T\neq$ 0, that is,
\begin{equation}
N = \sum_{k'}(2 -\delta_{k'k})f_{k'}~,
\label{N zero pairing}
\end{equation}
where $f_{k'}$ is the single-particle occupation number
\begin{equation}
f_{k'} = \frac{1}{e^{\beta(\epsilon_{k'} - \lambda)}+1}~.
\label{fk}
\end{equation}

\noindent Instead of this, the zero-pairing limit of Eq. (\ref{Nodd}) is $N = 1+ 2\sum_{k'\neq k}f_{k'}$.

The second shortcoming of the Maino's approach is that the imposed requirement (\ref{limR}) for the zero-temperature limit causes a jump between the average pairing gap as well as the total energy, obtained at $T=$ 0, and their corresponding values at finite $T$ because at $T=$ 0 the average gap (total energy) is equal to that, obtained from Eq. (\ref{oddBCSgap}) for  $k=k_0$, whereas at $T\neq$ 0 it is defined as the weighted sum over all the gaps (total energies), obtained from the same equation but with different $k$. There is no smooth transition between the two average gaps (total energies).

To remove these shortcomings, we propose to derive the equation for the odd particle number as follows. From Eq. (\ref{Neven}) for the even particle number, it is clear that the equation for the odd particle number with the odd particle occupying the level $k$ can be written as
\begin{equation}
N^{(k)} =\sum_{k'}(2-\delta_{k'k})[n_{k'} u^2_{k'}+ (1-n_{k'}) v_{k'}^2] 
= q_k + 2\sum_{k'\neq k}[n_{k'} u^2_{k'}+ (1-n_{k'}) v_{k'}^2]~,
\label{Nodd q} 
\end{equation} 
where
\begin{equation}
q_k = n_k u^2_k+ (1-n_k) v_k^2 = 1-n'_k~,\hspace{5mm} n'_k = \frac{1}{e^{\beta|\epsilon_k-\lambda|}+1}~,
\label{n'}
\end{equation}
because, for the unpaired particle on level $k$, one has $u_k=$ 0, $v_k=$ 1, and $n_k=n'_k$ as the pairing gap $\Delta$ = 0. 
In the zero-pairing limit, $q_k = f_k$ and the sum at the right-hand side of Eq. (\ref{Nodd q}) reduces to $2\sum_{k'\neq k}f_{k'}$. Hence
one recovers Eq. (\ref{N zero pairing}) for $N$ noninteracting single particles.
By neglecting $n'_k$, one obtains $q_k$ = 1, and recovers from Eq. (\ref{Nodd q}) the particle-number equation (\ref{Nodd}) in the Maino's approach.
We refer to Eqs. (\ref{oddBCSgap}) and (\ref{Nodd q}) as the ``$q_k$-blocked BCS".

Assuming $k=k_0$ in Eq. (\ref{Nodd q}), it follows that any other level $k \neq k_0$ in the system has the probability $1-q_{k_0}$ to be occupied by the odd particle. The average of any observable $\hat{\cal O}^{(k_0)}$ is then calculated as
\begin{equation}
\langle\hat{\cal O}^{(k_0)}\rangle = \frac{1}{\cal N}\bigg[q_{k_0}{\cal O}^{(k_0)}+(1-q_{k_0})\sum_{k\neq k_{0}}{\cal O}^{(k)}\bigg]~, \hspace{5mm}
{\cal N} ={q_{k_0} + (\Omega-1)(1-q_{k_0})}~
\label{Aveq}
\end{equation}
with $\Omega$ being the total number of levels in the system. In the zero-temperature limit, $q_{k_0} =$ 1 and the average (\ref{Aveq}) smoothly reduces to its ground-state value without imposing any special requirement at $T=$ 0. We refer to this approach, which calculates all the thermodynamic quantities by using the average (\ref{Aveq}) based on the $q_k$-blocked-BCS Eqs. (\ref{oddBCSgap}) and (\ref{Nodd q}), as the ``average-blocked BCS".

It is now well established that, in finite systems such as nuclei, thermal fluctuations smooth out the sharp phase transition from the superfluid phase to the normal one~\cite{Moretto2,MBCS,Zele}. It has been demonstrated in Refs. \cite{MBCS,FTBCS1}, by taking into account the effect owing to the QNF, which is neglected in the standard FTBCS, that the pairing gap does not collapse at $T_c$, but monotonically decreases with increasing $T$. In the present paper, we also include the effect of the QNF within the FTBCS1~\cite{FTBCS1}, whose result is the gap equation
\begin{equation}
\Delta^{(k)}_{k'} = G\bigg[\sum_{k''\neq k}(1-2n_{k''})u_{k''}v_{k''} 
+ 2(1-\delta_{kk'})\frac{n_{k'}(1-n_{k'})}{1-2n_{k'}}u_{k'}v_{k'}\bigg]~,
\label{oddBCS1gap}
\end{equation} 
from which one obtains the level-independent gap $\bar{\Delta}^{(k)}$ for the odd system with the odd particle occupying the level $k$. This level-independent gap is obtained by averaging the gaps (\ref{oddBCS1gap}) over all the levels $k'\neq k$, namely,
\begin{equation}
\bar{\Delta}^{(k)}=\frac{1}{\Omega-1}{\sum_{k'\neq k}{\Delta}^{(k)}_{k'}}~.
\label{bargap}
\end{equation}

The last term on the right-hand side of Eq. (\ref{oddBCS1gap}) contains the QNF $\delta{\cal N}_{k'}^2\equiv n_{k'}(1-n_k')$. We refer to this approximation based on Eq. (\ref{Nodd q}) and (\ref{bargap}) as the $q_k$-blocked BCS1.
   
\subsection{Finite-temperature extension of exact solutions for odd systems}
The exact solutions of the pairing Hamiltonian (\ref{Heven}) were first obtained by Richardson in 1963 by solving the Richardson's nonlinear equation for each total seniority quantum number $S$, which is the number of unpaired particles in the system~\cite{Exact}. Fifteen years ago, a method based on SU(2) algebra of the angular momentum, which diagonalizes directly the Hamiltonian (\ref{Heven}) instead of solving the Richardson's equations, was proposed in Ref. \cite{Exact1}. This method allows us to find simultaneously all the exact eigenvalues $\cal{E}_S^{\text{(ex)}}$ and exact occupation numbers $f_k^{S}$ with the total seniority $S = 0, 2, ..., \Omega$ for even systems. For the odd systems, since the odd particle does not participate in the pairing correlation, it is separated from the off-diagonal matrix elements of the pairing Hamiltonian, which describe the pair transfer (see Eq. (9) of Ref. \cite{Exact1}, for example). This separation is called the blocking effect caused by the odd particle in the exact pairing method. As the result, the exact ground state of the odd system corresponds to $S=1$, whereas the excited states correspond to $S = 3, 5, ..., \Omega$. The exact ground-state energy is obtained when the odd particle occupies the highest level, which is located just above the Fermi surface, that is the level $k = k_0$. The extension of the exact solutions to finite-temperature is often carried out by using the CE, whose partition function is constructed based on the exact eigenvalues ${\cal E}_S^{\text{(ex)}}$ in the form~\cite{Ensemble}
\begin{equation}
Z^{\text{(ex)}} = \sum_S d_S e^{-{\cal E}_S^{\text{(ex)}}/T} ~, 
\label{ZCE}
\end{equation}
where $d_S = 2^S$ is the degeneracy. Knowing the partition function, one can easily obtain all the exact thermodynamic quantities of the system such as the total energy ${E}^{\text{(ex)}}$, entropy $S^{\text{(ex)}}$, and heat capacity $C^{\text{(ex)}}$, namely, 
\begin{equation}
E^{\text{(ex)}} = \frac{T^2}{Z^{\text{(ex)}}}\frac{\partial\text{ln}Z^{\text{(ex)}}}{\partial T} ~, \hspace{5mm}
S^{\text{(ex)}} = \beta{E}^{\text{(ex)}} + \text{ln}Z^{\text{(ex)}} ~, \hspace{5mm}
C^{\text{(ex)}} = \frac{\partial{E}^{\text{(ex)}}}{\partial T} ~.
\label{Exact_TQs}
\end{equation}

The exact temperature-dependent occupation numbers $f_k^{\text{(ex)}}$ are obtained by averaging the exact state-dependent occupation numbers $f_k^{(S)}$ over the CE, namely,
\begin{equation}
f_k^{\text{(ex)}} = \frac{1}{Z^{\text{(ex)}}(T)}\sum_S d_S f_k^{(S)} e^{-{\cal E}_S^{\text{(ex)}}/T} ~. 
\label{fk_Exact}
\end{equation}

Based on the exact total energy $E^{\text{(ex)}}$ and occupation numbers $f_k^{\text{(ex)}}$, one can compute the exact pairing gap from the relation of the mean-field (BCS) gap as~\cite{Ensemble} 
\begin{equation}
\Delta^{\text{(ex)}} = \sqrt{-G\bigg\{E^{\text{(ex)}}-2\sum_k \epsilon_k f_k^{\text{(ex)}} + G\sum_k [f_k^{\text{(ex)}}]^2 \bigg\}} ~.
\label{Gap_Exact}
\end{equation}

\section{Analysis of numerical results}
\label{results}
For the clarity of comparison between the Maino's and our approaches,  we employed the schematic model, which consists of $N$ particles ($N$ is odd) occupying $\Omega$ doubly folded levels ($N\leq\Omega$) and interacting via a pairing force with the strength parameter $G$. Various values of $N$, $\Omega$, and $G$ are tested by using equidistant and nonequidistant levels. In the discussion below we will analyze the most representative cases, namely with $N=$ 9, $G = $ 0.6 MeV, $\Omega=$ 10, 14, and 20, with the level distant equal to 1 MeV, that is, with the  single-particle energies (in MeV) $\epsilon_k = k$ ($k= 1, 2,...,\Omega$). An example is also shown for the non-equidistant case with $\Omega=$ 10 by using the following values of $\epsilon_k$: $\epsilon_1=$ 1, $\epsilon_2=$ 2, $\epsilon_3=$ 2.5, $\epsilon_4=$ 4, $\epsilon_5=$ 5.5,  $\epsilon_6=$ 6.2, $\epsilon_7=$ 6.8, $\epsilon_8=$ 7, $\epsilon_9=$ 7.2, $\epsilon_{10}=$ 7.5.

\begin{figure}[h]
\begin{center}
\includegraphics[width=10cm]{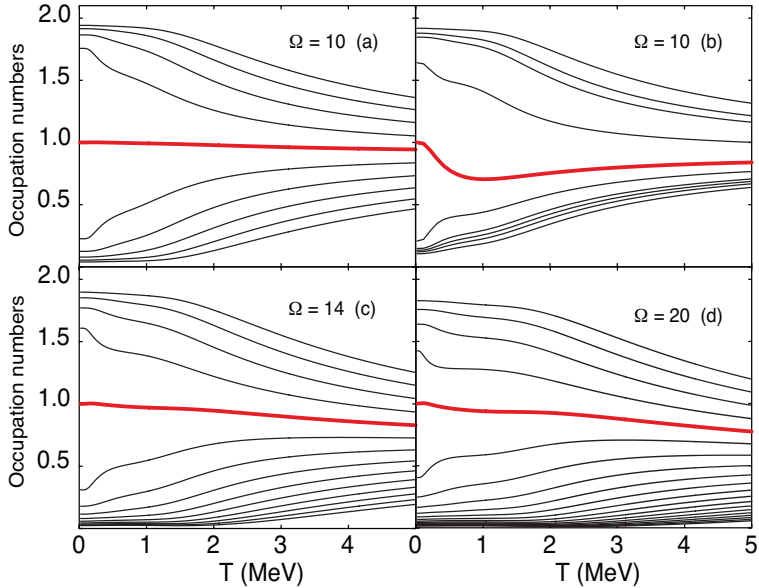}
\caption{(Color online) Exact occupation numbers of the levels as functions of $T$ for $N=$ 9, $\Omega=$ 10, 14, and 20. (a), (c), and (d) show the results obtained with equidistant levels; (b) - those with nonequidistant ones (see text). In each panel, the higher-located line is the occupation number of the lower-located level. The thick line, which starts from 1 at $T$ = 0,  is the occupation number of the level occupied by the odd particle. \label{exact f}}
\end{center}
\end{figure}

Shown in Fig. \ref{exact f} are the exact occupation numbers of the levels as functions of temperature $T$ for $\Omega=$ 10, 14, and 20. It is seen from this figure that, at $T=$ 0, pairing decreases (increases) the occupation numbers of the levels below (above) the Fermi surface from 2 (0), which is their values in the absence of pairing. With increasing $T$, the occupation numbers of the levels below (above) the Fermi surface smoothly decrease (increase). The results shown in this figure reveal two important features. First, despite the odd particle is allowed to occupy any level in diagonalizing the pairing Hamiltonian, in average, it always occupies the fifth level, which is the highest occupied level ($k_0=$ 5 for $N=$ 9) in the ground state ($T=$ 0), because this is the only level with the occupation number equal to 1  ($f_{k_0} =$ 1) at $T=$ 0. This feature rules out the assumption that one has to average $f_k$ over all the levels above the chemical potential as in the Maino's approach. In fact, the odd particle occupies only the level $k_0$. Second, the occupation number of the odd level $f_{k_0}$ does not remain equal to 1, but decreases with increasing $T$ in the equidistant model for all values of the level number $\Omega$ under consideration [Figs. \ref{exact f}(a), \ref{exact f}(c), and \ref{exact f}(d)].  In the nonequidistant model, $f_{k_{0}}$ decreases significantly as $T$ increases from 0 to 1 MeV, then increases slightly as $T$ increases further, but still remaining smaller than 1 up to $T=$ 5 MeV. This feature is a clear evidence that the assumption $q_{k_0}=$ 1 as in the blocked-BCS Eq. (\ref{Nodd}) might not be correct.

\begin{figure}[h]
\begin{center}
\includegraphics[width=9cm]{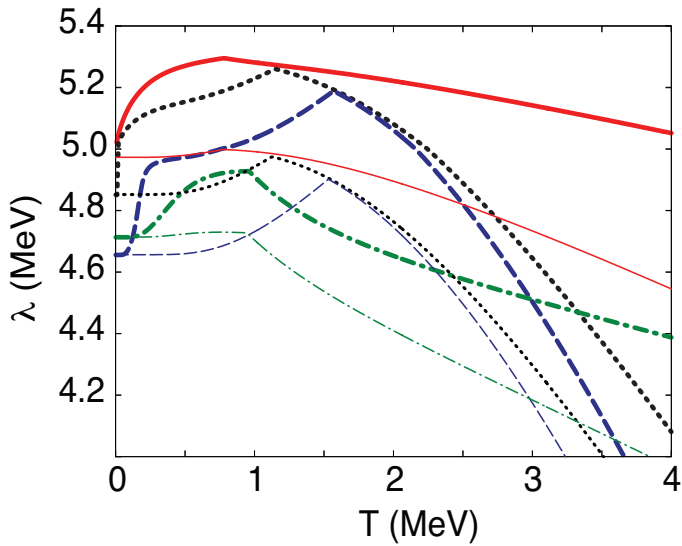}
\caption{Chemical potentials as functions of $T$ for $N=$ 9. The solid, dotted, and dashed lines are results obtained within the equidistant model with $\Omega$ = 10, 14, and 20, respectively. The dot-dashed line stands for the prediction obtained within the nonequidistant model with $\Omega=$ 10.  The thin lines correspond to the predictions by the blocked BCS, whereas the thick lines are those obtained within the $q_k$-blocked BCS.
\label{lambda}}
\end{center}
\end{figure}
The chemical potentials obtained within the $q_k$-blocked BCS (with $k=k_0$), and blocked BCS, that is with $q_{k_0} =$ 1, are shown in Fig. \ref{lambda} as functions of $T$ for $\Omega=$ 10, 14, and 20. The chemical potentials increase with $T$ up to $T=T_c$, where, because of the collapse of the pairing gap, their temperature dependence abruptly changes to decreasing as $T$ increases further. The figure is a clear demonstration of the large difference between the two approximations. Indeed, while $q_k$ within the $q_k$-blocked BCS varies with $T$ as in Eq. (\ref{n'}), it remains always 1 independently of $T$ within the blocked BCS. As the result the chemical potentials $\lambda$ obtained as the solutions of Eqs. (\ref{Nodd}) and (\ref{Nodd q}) coincide only at $T=$ 0, where $q_{k_0} =$ 1 in both approximations. Consequently, the chemical potential can become even larger than the energy $\epsilon_{k_0}$ (5 MeV) of the blocked level $k_0$ within the $q_k$-blocked BCS (thick lines), whereas it remains always below the blocked level within the blocked BCS (thin lines).

\begin{figure}[h]
\begin{center}
\includegraphics[width=16cm]{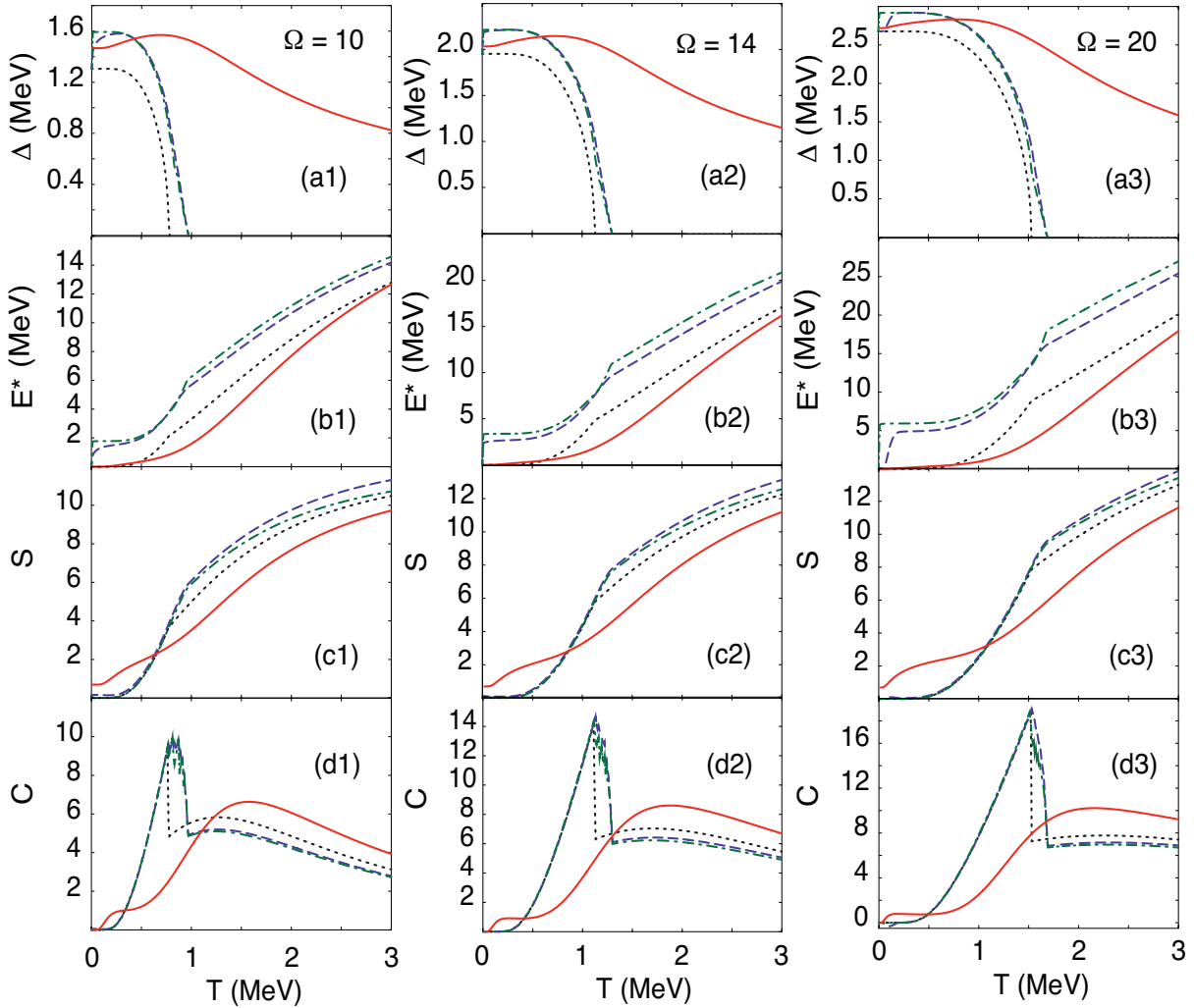}
\caption{(Color online) Pairing gap $\Delta$ [(a1)--(a3)], excitation energy $E^{*}$ [(b1)--(b3)], entropy $S$ [(c1)--(c3)], and heat capacity $C$ [(d1)--(d3)] as functions of $T$ for the equidistant model with $N=$ 9, $\Omega=$ 10, 14, and 20. The dotted, dashed, dot-dashed, and solid lines stand for the predictions obtained by using the blocked BCS, average-blocked BCS, Maino's approach, and exact solutions, respectively. 
\label{FTBCS}}
\end{center}
\end{figure}
The pairing gap, excitation energy, entropy, and heat capacity, which are predicted by the $q_k$-blocked BCS, average-blocked BCS, Maino's approach, and exact solutions are shown in Fig. \ref{FTBCS} as functions of $T$ for the equidistant model with $\Omega=$ 10, 14, and 20. The figure shows that the pairing gap and the heat capacity obtained within the average-blocked BCS reproduce remarkably well the predictions by the Maino's approach. The excitation energy and entropy predicted by the average-blocked BCS are slightly lower than those obtained within the Maino's approach, and therefore, closer to the blocked-BCS results. The latter are closest to the results obtained by using exact solutions.

\begin{figure}[h]
\begin{center}
\includegraphics[width=10cm]{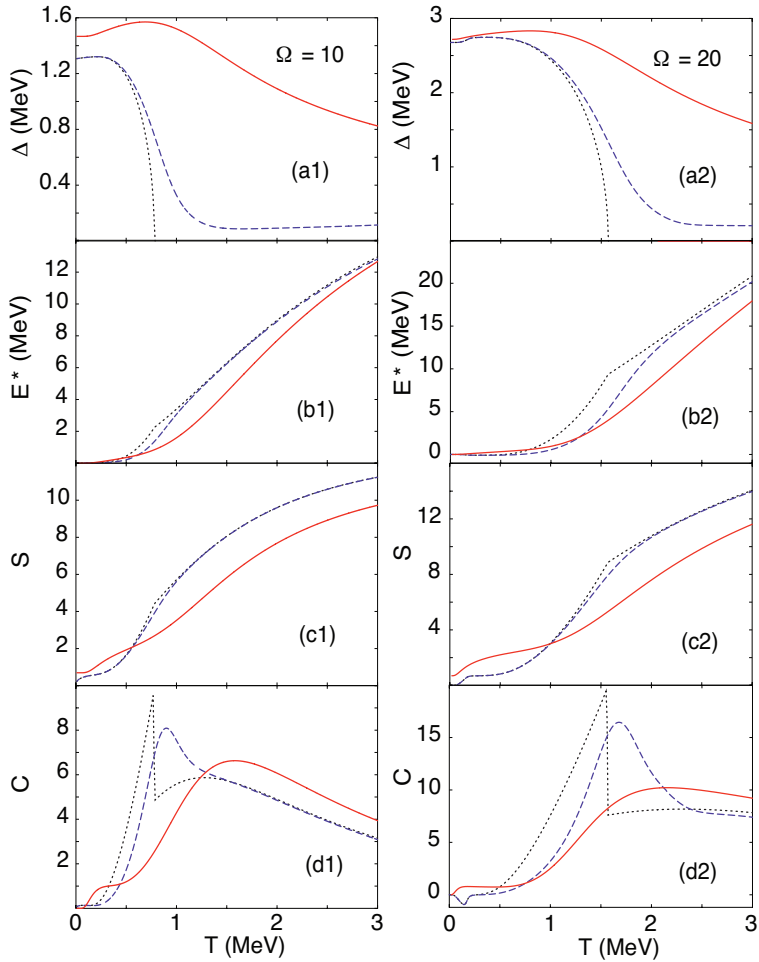}
\caption{(Color online) Same thermodynamic quantities as in Fig. \ref{FTBCS} for the equidistant model with $N=$ 9, $\Omega=$ 10 and 20. The dotted, dashed, and solid lines denote the predictions obtained by using the $q_k$-blocked BCS, $q_k$-blocked BCS1, and exact solutions, respectively.
\label{FTBCS1}}
\end{center}
\end{figure}
The same thermodynamic quantities obtained within the $q_k$-blocked BCS are compared with the predictions by the $q_k$-blocked BCS1 and the exact solutions for $\Omega=$ 10 and 20 in Fig. \ref{FTBCS1}. The sharp superfluid-normal phase transition in the $q_k$-blocked BCS (dotted lines) is smoothed out in the $q_k$-blocked BCS1 by taking the QNF into account. As compared to the blocked BCS, where $q_k=$ 1, the pairing gap obtained within the $q_k$-blocked BCS slightly increases with $T$ at $T<$ 0.5 MeV because of the blocking effect becomes weaker, whereas $T$ is not sufficiently high to break the first pair. The similar trend is seen in the exact pairing gap, where it keeps increasing up to $T\sim$ 0.7--0.8 MeV. This increase in the pairing gap leads to a slight decrease  in the excitation energy at low $T$ for $\Omega=$ 20. Consequently, the heat capacity turns negative at a very low $T\simeq$ 0.15 MeV [Fig. \ref{FTBCS1} (d2)]. This negative heat capacity is due to the equidistant levels under consideration. A test by using the modified spectrum with $\epsilon_5=$ 4.3, $\epsilon_6=$ 5.8, and $\epsilon_7=$ 7.5 MeV shows no negative value of the heat capacity.

Because the exact single-particle occupations numbers $f_{k \neq k_0}^{\text{(ex)}}$ are neither 0 nor 1 (Fig. \ref{exact f}), the exact entropies are finite whereas the quasiparticle entropies are zero at $T=$ 0. Since the $q_k$-blocked BCS1 is just a correction of the $q_k$-blocked BCS by including the effect of QNF, neglected within the standard BCS, it cannot improve much the BCS except for the smoothing out of the superfluid-normal phase transition. Moreover, particle number projection has yet to be done. Therefore there is still a large discrepancy between the predictions by the $q_k$-blocked BCS1 and those obtained from the exact solutions. But at least, on the qualitative level, it shows the right trend towards the exact results.
\section{Conclusions}
In the present work we propose an improved treatment of the blocking effect in the systems with odd numbers of fermions interacting via the constant pairing force. By using the exact solutions of the pairing Hamiltonian, we show that the conventional assumption of the unity occupation number for the blocked level occupied by the odd particle is not valid at $T\neq$ 0. Instead, we introduce the temperature-dependent occupation number $q_k$ for the blocked level, which is directly derived from the standard equation for the particle number within the standard FTBCS. We also construct the average $q_k$-blocked BCS, which reproduces quite well the predictions by the Maino's approach. However, again by using the exact solutions of the pairing Hamiltonian, we demonstrate that the average procedure over all the levels above the Fermi surface as proposed in the Maino's approach does not correspond to the real situation observed in the exact solutions, where the odd particle actually always remains on the top level $k=k_0$, which is located above the Fermi surface at $T=$ 0.

Based on the analysis of the results obtained in the present paper, we believe that, in the study of the odd systems with pairing, such as atomic nuclei, at finite temperature, if the BCS approach with blocking is ever applied, the $q_k$-blocked BCS with $k=k_0$ proposed here should be used instead of the blocked BCS, whereas neither the Maino's approach nor the average $q_k$-blocked BCS is necessary.
\acknowledgments
The numerical calculations were carried out using the \footnotesize FORTRAN \normalsize IMSL
Library by Visual Numerics on the RIKEN supercomputer HOKUSAI-GreatWave  System. N.Q.H. acknowledges the support by the National Foundation for Science and Technology Development(NAFOSTED) of Vietnam through Grant No. 103.04-2013.08. 

\end{document}